\documentclass[a4paper,fleqn]{cas-dc}

\usepackage[authoryear,longnamesfirst]{natbib}
\usepackage[authoryear]{natbib}

\def\tsc#1{\csdef{#1}{\textsc{\lowercase{#1}}\xspace}}
\tsc{WGM}
\tsc{QE}
\tsc{EP}
\tsc{PMS}
\tsc{BEC}
\tsc{DE}

\begin{document}
\let\WriteBookmarks\relax
\def\floatpagepagefraction{1}
\def\textpagefraction{.001}
\shorttitle{Multiple regression analysis of 
anthropogenic and heliogenic climate drivers, and some cautious forecasts}
\shortauthors{Frank Stefani}

\title [mode = title]{Multiple regression analysis of 
anthropogenic and heliogenic climate drivers, and some 
cautious forecasts}

\author[1]{Frank Stefani}[orcid=0000-0002-8770-4080]
\ead{F.Stefani@hzdr.de}
\address[1]{Helmholtz-Zentrum Dresden-Rossendorf, 
Institute of Fluid Dynamics,
Bautzner Landstr. 400, 01328 
Dresden, Germany}

\begin{abstract}
The two main drivers of climate change on sub-Milankovic 
time scales are re-assessed by means of a multiple regression
analysis. Evaluating linear combinations of 
the logarithm of carbon dioxide concentration 
and the geomagnetic aa-index as a proxy for solar activity, we 
reproduce the sea surface temperature (HadSST)
since the middle of the 19th century 
with an adjusted $R^2$ value of around 87 per cent 
for a climate sensitivity (of TCR type) in the range of 
0.6\,K until 1.6\,K per doubling of CO$_2$.
The solution of the regression is quite sensitive: 
when including data from the last decade, the simultaneous 
occurrence of a strong El Ni{\~n}o on one side and 
low aa-values on the other
side lead to a preponderance of solutions with 
relatively high climate sensitivities around 1.6\,K. 
If those later data are excluded, the regression leads to 
a significantly 
higher weight of the aa-index and a correspondingly 
lower climate sensitivity going down to 0.6\,K. 
The plausibility of such low values is 
discussed in view of recent experimental and 
satellite-borne 
measurements. We argue that a further decade of 
data collection will be needed to allow for
a reliable distinction between low and high sensitivity values.
Based on recent ideas about a quasi-deterministic 
planetary synchronization of the solar dynamo, we make 
a first attempt to predict the aa-index and 
the resulting temperature anomaly for various 
typical CO$_2$ scenarios.
Even for the highest climate sensitivities, and an
unabated linear CO$_2$ increase, we predict 
only a mild additional temperature rise of around 1\,K 
until the end of the century, while for the lower 
values an imminent temperature drop 
in the near future, followed by a rather flat temperature 
curve, is prognosticated.

\end{abstract}

\begin{keywords}
Climate change \sep Solar cycle \sep Forecast
\end{keywords}

\maketitle

\section{Introduction}
As heir of its great pioneers  \cite{Arrhenius1906}
and  \cite{Callendar1938}, modern climate science
\citep{Knutti2017}
has  been surprisingly unsuccessful 
in narrowing down
its most prominent parameter -  
equilibrium climate sensitivity (ECS) - from the 
ample range 1.5\,K-4.5\,K (per $2\times$\,CO$_2$) 
as already given in the report by  
\cite{Charney1979}. 
This sobering scientific yield is often discussed 
in terms of various interfering socio-scientific 
and political factors
\citep{Hart2015,Lindzen2020,Vahrenholt2020}.
Yet, in addition to those more ``subjective'' 
reasons for climate science to be that {\it unsettled},
there are at least two ``objective'' ones: 
the lack of precise and reliable  experimental 
measurements of the climate sensitivity until 
very recently, and the unsatisfying state of
understanding the complementary solar influence on the 
climate. Certainly, a couple of mechanisms have been proposed
\citep{Hoyt1993,Gray2010,Lean2010} that could significantly 
surmount the 
meager 0.1 per cent variation of the total solar irradiance 
(TSI) which is routinely used as an argument against any 
discernible solar impact on the climate. Among those mechanisms,
the following ones figure most prominently: 
the comparable large variation of the UV component with  
its influence on the ozone layer and the resulting 
stratospheric-tropospheric 
coupling \citep{Labitzke1988,Haigh1994,Soon2000,Georgieva2012,Silverman2018,Veretenenko2020}; 
the effects of solar magnetic field modulated cosmic rays
on aerosols and clouds \citep{Svensmark1997,Soon2000b,Shaviv2003,Svensmark2017};
downward winds following geomagnetic storms in the polar caps of the
thermosphere, penetrating stratosphere and troposphere \citep{Bucha1998};  
solar wind's impact on the global electric current \citep{Tinsley2000,Tinsley2008};
and the (UV) radiation effects on the growth of oceanic phytoplankton 
\citep{Vos2004} which, in turn, produces dimethylsulphide,
a major source of cloud-condensation nuclei 
\citep{Charlson1987}. But even the very TSI 
was claimed \citep{Hoyt1993,Scafetta2014,Egorova2018,Connolly2021}
to have risen much more steeply 
since the Little Ice Age than assumed in
the conservative estimations by
\cite{Wang2005,Steinhilber2009,Krivova2010}.
While neither of those mechanism can presently 
be considered as conclusively proven 
\citep{Solanki2002,Courtillot2007,Gray2010}, 
they all together entail significantly more potential for 
solar influence on the terrestrial climate than what was   
discussed on the corresponding one and a half pages 
of \cite{Bindoff2013}.

In view of illusionary claims \citep{Cook2016}
of an overwhelming scientific 
consensus on this complex and vividly debated 
research topic, and the severe 
political consequences drawn from it, 
we reiterate here Eugene Parker's prophetic warning 
\citep{Parker1999}  that ``...it is essential 
to check to what extend the facts support these
conclusions before embarking on drastic, perilous
and perhaps misguided plans for global action''.
We also agree with his ``...inescapable conclusion (...) 
that we will have to know a lot more about the sun 
and the terrestrial atmosphere 
before we can understand the nature of contemporary changes 
in climate''.

Thus motivated, and also provoked by  
recent experimental \citep{Laubereau2013} 
and satellite-borne measurements   
\citep{Feldman2015,Rentsch2019} which 
pointed consistently to a rather
low climate sensitivity, we make here another 
attempt to quantify the respective shares of anthropogenic 
and heliogenic climate drivers.
Specifically, we resume the long tradition of
correlating terrestrial temperature data with 
certain proxies of  solar activity, as pioneered by
\cite{Reid1987} for the sunspot numbers, 
by \cite{Friis1991,Solheim2012} for the solar cycle length, 
and by \cite{Cliver1998,Mufti2011} for the geomagnetic 
aa-index \citep{Mayaud1972}. 
Our work builds strongly
on the two latter papers, which - based on 
data ending in 1990 and 2007, respectively - 
had found empirical correlation coefficients
between the aa-index and temperature variations of 
up to 0.95.
Notwithstanding some doubts regarding their 
statistical validity \citep{Love2011},
such remarkably high correlations might rise the 
provocative
question of whether any sort of greenhouse effect {\it is
still needed at all} to explain the (undisputed) global warming
over the last one and a half century. 
More recently, however, any prospects for such a 
{\it reversed} simplification
were dimmed by the fact that the latest decline 
of the aa-index was not accompanied by a 
corresponding drop of temperature. By contrast, the latter 
remained rather constant during the first one and a 
half decades of the 21st century (the ``hiatus''), 
and even increased with the recent strong 
El Ni{\~n}o events. 

This paper aims at supplementing the previous work of 
\cite{Cliver1998,Pulkkinen2001,Mufti2011,Zherebtsov2019}
by taking seriously into account both observations: the nearly 
perfect correlation of solar activity with temperature over 
about 150 years, and the notable 
divergence between those quantities during the last two 
decades.
Using a multiple regression analysis,
quite similar to that of \cite{Soon1996}, but 
with the time series of the aa-index as the 
second independent variable (in addition to the 
logarithm of CO$_2$ concentration),
we will show that the temperature variation
since the middle of the 19th century
can be reproduced  with an (adjusted) $R^2$ value 
around 87 per cent.
Such a goodness-of-fit is  
achieved for specific combinations of the weights of 
the aa-index and of CO$_2$ that form a nearly linear function
in their two-dimensional parameter space.
Best results are obtained for a climate sensitivity 
in the range between 0.6\,K-1.6\,K (per 2$\times$ CO$_2$), 
with a delicate dependence 
on whether the latest data are included or 
not. Derived from empirical variations on the
(multi-)decadal time scale, this climate sensitivity
should be interpreted as a transient climate response 
(TCR), rather than an ECS. 
Our range 
corresponds well with that 
of \cite{Lewis2018}, 0.8\,K-1.3\,K, 
but is appreciably
lower than the ``official'' 1.0\,K-2.5\,K
range \citep{Knutti2017}. 
The lower edge of our estimation will be plausibilized
by recent experimental \citep{Laubereau2013} and 
satellite-borne measurements \citep{Feldman2015,Rentsch2019}.
It is also quite close, although still higher, then the 
particularly low estimate of less than 0.44\,K,
as advocated by  \cite{Soon2015} after comparing 
exclusively rural temperature data in the Northern hemisphere 
with the TSI. The upper edge, in turn, is not far from the  
spectroscopy-based estimation by \cite{Happer2020}.

With the complementary share of the Sun for global warming 
thus reaching values between 30 and 70 per cent, any climate 
forecast will require a descent prediction of solar activity. 
This leads us into yet another
controversial playing field, viz, the predictability
of the solar dynamo. While the existence of the short-term 
Schwabe/Hale cycles is a truism in the solar 
physics community, the existence and/or stability of the 
mid-term Gleissberg and Suess-de Vries 
cycles are already controversially discussed, and there
is even more uncertainty about long-term variations such as  
the Eddy and  Hallstatt ``cycles'', which are closely related 
to the sequence of Bond events \citep{Bond2001}.

In a series of recent papers 
\citep{Stefani2016,Stefani2017,Stefani2018,Stefani2019,Stefani2020a,Stefani2020b,Stefani2020c}, 
we  have tried to develop a self-consistent explanation of those short-, 
medium- and long-term  solar cycles in terms of synchronization by 
planetary motions. According to our present understanding, the surprisingly 
phase-stable 22.14-year Hale cycle \citep{Vos2004,Stefani2020b}
results from parametric resonance of a 
conventional $\alpha-\Omega$ dynamo with an 
oscillatory part of the helical 
turbulence parameter $\alpha$ that
is thought to be synchronized by the 11.07-year spring-tide periodicity 
of the three tidally dominant planets Venus, Earth, and Jupiter
\citep{Stefani2016,Stefani2017,Stefani2018,Stefani2019}.
The medium-term Suess-de Vries cycle (specified to 
193 years in our model) emerges then as a {\it beat period}
between the basic 22.14-year Hale cycle and some (yet not well understood)
spin-orbit coupling connected with the 
motion of the Sun around the barycenter of
the solar system that is governed by the 19.86-year synodes of
Jupiter and Saturn \citep{Stefani2020a,Stefani2020c}. Closely related to this, 
some Gleissberg-type 
cycles appear as nonlinear beat effects and/or 
from perturbations of Sun's orbital motion from other synodes 
of the Jovian planets.  
Finally, the  long-term 
variations on the millennial time-scale (Bond events) 
arise as chaotic  transitions 
between regular and irregular episodes of the solar dynamo  
\citep{Stefani2020c}, in close analogy with the super-modulation 
concept introduced by \cite{Weiss2016}.

Being well aware of the conjectural nature of this synchronized solar 
dynamo model, we nevertheless dare to make a cautious prediction of 
the aa-index for the next 130 years, based on some simple 3-frequency 
fits to the aa-index data over the last 170 years. While
our choice for three frequencies is motivated by the dominance of 
one Suess-de Vries and two Gleissberg-type cycles, we employ
different versions of fixing or relaxing their frequencies, which 
leads to a certain variety of forecasts. A common feature of all of 
them is, however, a noticeable decline of solar activity 
until 2100, and a recovery in the 22nd century. Those predictions 
for the aa-index are then combined with three different
scenarios of CO$_2$ increase, including an unfettered annual 
increase by 2.5 ppm and two further scenarios based 
on hypothetical decarbonization schemes.
The $3 \times 3$ models thus obtained are then blended with 
the different combinations of weights for the aa-index and 
CO$_2$ as derived before in the regression analysis. For  
the ``hottest'' scenario 
we predict an additional temperature increase until 2100 
of less than $1$\,K, 
while all other combinations lead to less warming, partly 
even to some imminent cooling, 
followed by a rather flat behaviour in which decreasing  
solar activity and a mildly increasing trend from CO$_2$ 
compensate each other to a large extend.

\section{Multiple regression analysis}

In this section, we perform  a multiple (or better: double) 
regression analysis
of the temperature data (dependent data) on the geomagnetic 
aa-index and the logarithm of the CO$_2$ (independent data). 
We do this in an intuitive and easily reproducible way 
by showing the  {\it fraction of variance unexplained (${\rm FVU}$)}, 
i.e. the ratio of the residual sum of squares to the 
total sum of squares, whose minimum is then identified. 
From those ${\rm FVU}$'s we will derive the corresponding 
$R^2$ value, both in its usual and in its adjusted variant. Let 
us start, however, with a description of our data base.

\subsection{Data}

While reliable CO$_2$ data are available for quite a long time, 
we decided to restrict our data base to the time
from the middle of the 19th century, for which both 
temperature data and the aa-index are readily available.

\begin{figure}[t]		
\includegraphics[width=0.99\linewidth]{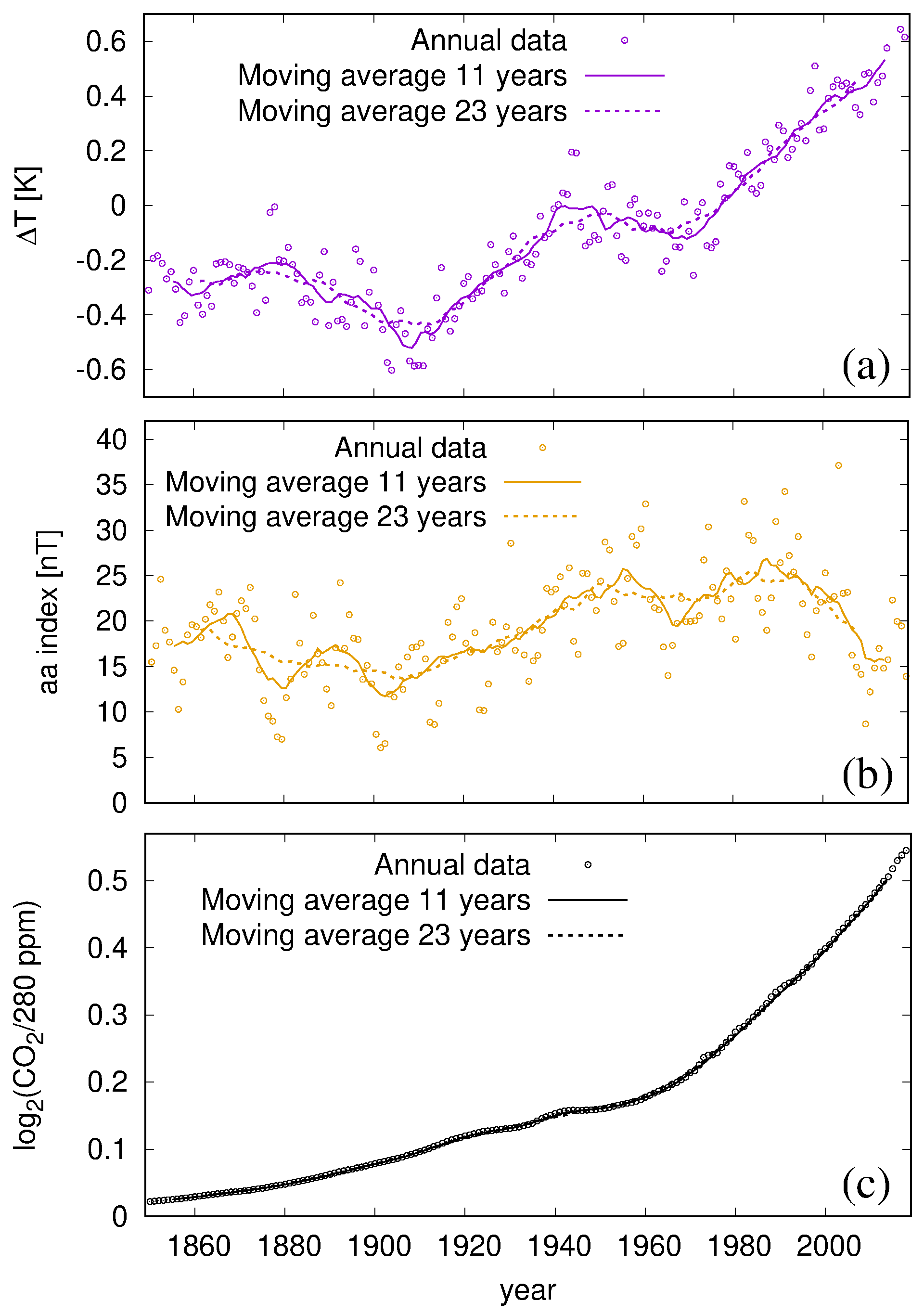}
\caption{Data of the HadSST sea surface temperature anomaly $\Delta T$ (a),
the aa-index (b), and $\log_2$ of the ratio of the CO$_2$ 
concentration to the reference value of 280 ppm (c). The annual 
data between 1850-2018 are 
complemented by centered moving averages with windows 
11 years and 23 years, as also utilized by \cite{Mufti2011}.
The sources of the data are described in the text.}
\label{FIG:1}
\end{figure}

\subsubsection{Temperature data}

There are quite a number of different temperature series which 
might be considered as basis for regression analysis. 
In order to make contact with the work of 
\cite{Mufti2011}, we decided to use the Hadley Centre Sea Surface 
Temperature (HadSST) data set in its updated version  
HadSST.4.0.0.0, available from 
www.metoffice.gov.uk, 
which provides us the sea surface temperature anomaly from 
1850 until 2018, relative to the 1961-1990 average
\citep{Kennedy2019}.
Actually, these data are not gravely different from the 
combined sea/land surface temperature (HadCRUT), apart from
some slight but systematic divergence during the last two 
decades.
At this point, our preference for HadSST is also supported by 
their better agreement with the UAH satellite  data
(starting only in 1978)
with their significantly broader spatial coverage.

The HadSST data are shown as open circles in Fig. 1a, together with
two exemplary  centered moving averages with windows 
11 years (full line) and 23 years (dashed line), which we 
will frequently refer to in this paper. These curves show the 
typical temporal structure comprising a slow decay between 1850 and 1905,  
a rather steep rise 
between 1905 until 1940, again a mild decay until 1970,
followed by a steep increase until 1998. As for the 
last two decades, we first see the ``hiatus'' 
between 1999 and 2014, being then  overwhelmed by 
the recent strong El Ni\~no events. We will come back to 
those latest years further below.

\subsubsection{aa-index data}
The aa-index measures the 
amplitude of global geomagnetic activity during 
3-hour intervals at two antipodal 
magnetic observatories, 
normalized to geomagnetic latitude $\pm 50^{\circ}$ 
\citep{Mayaud1972}.
Inspired by the work of \cite{Cliver1998} and \cite{Mufti2011} 
who pointed out the remarkable correlation of up to 0.95 
between (time averaged) temperature anomalies and the geomagnetic 
aa-index, we will use 
the latter data as a proxy for solar activity.
A viable alternative would have been to use sunspot data,
or various versions of the TSI, as exemplified by \cite{Soon2015}.
Given, on one side, their generally high correlation with sunspots 
numbers \citep{Cliver1998a}, and, on the other side, their
high reliability based on precise measurement down
to 1844 (which avoids some ambiguities concerning the
correct variability of the TSI \citep{Soon2015,Connolly2021}), 
we focus here exclusively on the aa-index, leaving multiple
regressions analyses with other solar data to future work.

The bulk of the aa-index data, between 1868-2010, was obtained 
from ftp.ngdc.noaa.gov.
As in \cite{Mufti2011}, the early segment between 1844 and 1867 was 
taken from \cite{Nevanlinna1993}. The latest segment,
between 2011 and 2018, was obtained from
www.geomag.bgs.ac.uk.
All these aa-index data were annually averaged. Together with their 
11-year and 23-year moving averages, the annual ${\rm aa}$ data are 
shown in Fig 1b. Already by visual inspection, between 1850 and 1990 
we observe a remarkable 
similarity of their shape with that of temperature, while 
after 1995 the aa-index steeply declines, whereas the temperature 
continues to increase. We will have more to say on that 
divergence further below.

\subsubsection{CO$_2$ data}

The CO$_2$ concentration data until 2014 were obtained from 
iac.ethz.ch/CMIP6/. The four additional data points from 2015 and 2018
were taken from the website www.co2.earth.
Together with their 11-year and 23-year moving averages the logarithm 
of those data is shown on Fig. 1c.

\subsection{Are we dealing with the most relevant data?}

After having presented the data that actually 
will be used in the remainder of 
this paper, we make a short break to consider whether these are indeed
the most relevant data. 
Our reliance on the aa-index might indeed be 
questioned, as other studies \citep{Vahrenholt2020} 
have claimed a strong
temperature dependence on  ocean-atmosphere variations, such as the 
Pacific Decadal Oscillation (PDO) \citep{Mantua2002}
and the Atlantic Multidecadal Oscillation
(AMO) \citep{Wyatt2012}, with their similar time structures
governed by a sort of 60-70-year ``cyclicity''. 
On the other hand, there is also evidence for 
direct correlations of the aa-index with regional features, 
such as the Northern Annular Mode (NAM) \citep{Roy2016}.
In order to make contact with those possible links, 
in Fig. 2 we show exemplarily the 23-year 
averages of the AMO and the PDO data, 
together with the previously shown 
aa-index (appropriately shifted and scaled) and $\Delta T$.
It is clearly seen that the aa-index and $\Delta T$ have a  
particularly parallel behaviour 
until 1990, say. There is also some similarity with AMO, while PDO 
has a different time dependence.

\begin{figure}[t]		
\includegraphics[width=0.99\linewidth]{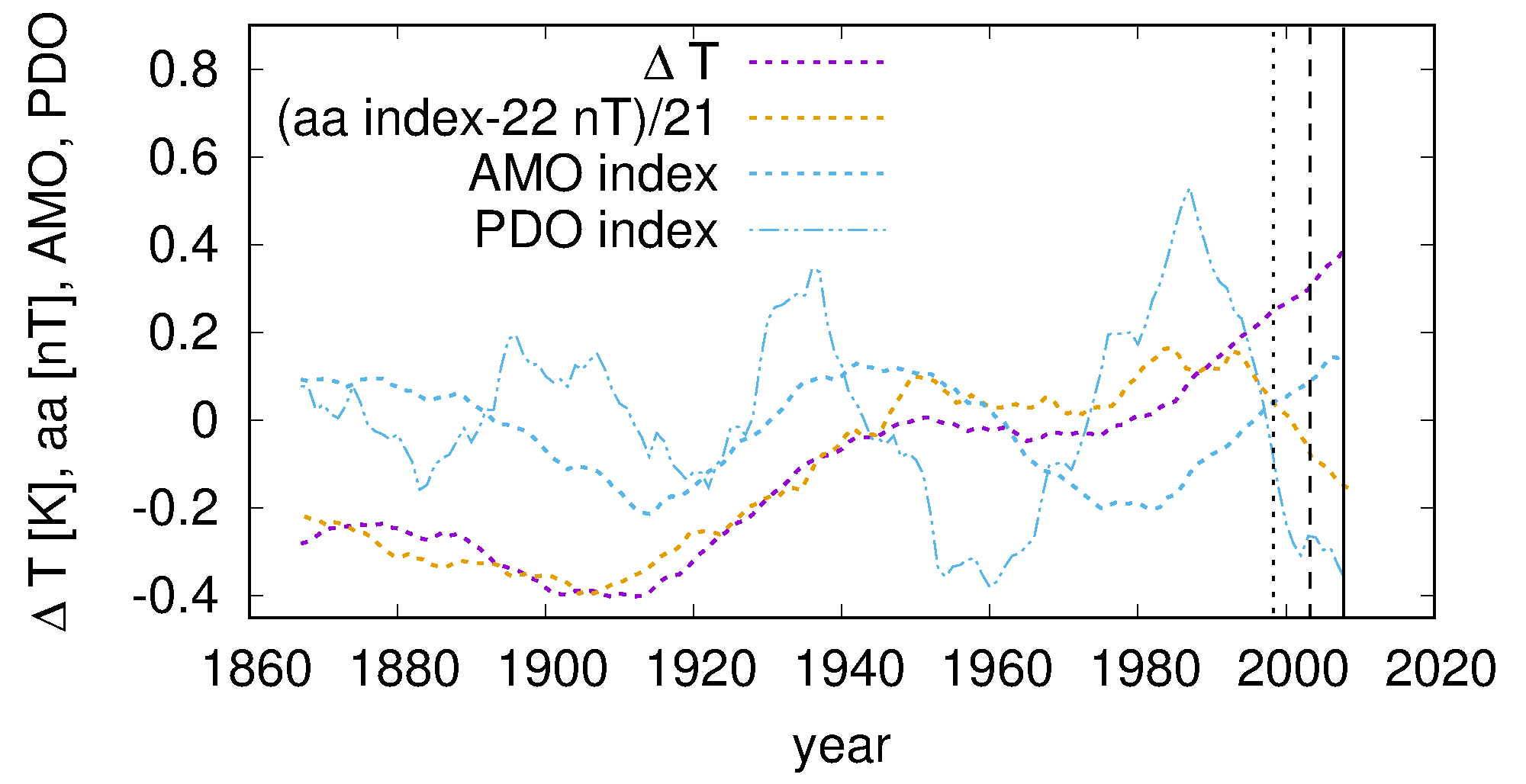}
\caption{Comparison between the centered moving averages over 23 years 
of the four data sets $\Delta T$, aa-index, AMO-index and PDO-index. 
Note the remarkable parallelity of $\Delta T$ and aa-index until 1990,
and the divergence thereafter. The AMO-index has also some similarity with 
$\Delta T$, while the PDO index is significantly different.}
\label{Fig:fig2}
\end{figure}

In Table 1 we quantify those relationships in terms of the 
empirical correlation coefficients $r$ for different 
data combinations. We do so for different end points of the time 
interval, namely 2008, 2003 and 1998 (note that these are the 
centered points of the last moving average interval, into which 
the annual data from up to 11 years later are included).
Our first observation is that both the correlations of CO$_2$ and 
of the aa-index with $\Delta T$ have similar $r$-values in the 
order of 0.9, which is a first indication for 
their comparable influences.
However, there are some subtleties to discern:
the correlation for CO$_2$ acquires its highest value ($r=0.916$) for 
the full time interval, and decreases slightly to 0.869 when the
time interval is shortened by 10 years.
By contrast, the correlation of aa-index with $\Delta T$ has only a value
$r=0.8$ for the full interval, but grows to  
0.95 for the restricted interval. This latter result confirms that 
of \cite{Mufti2011} obtained for a similar period, and by
\cite{Cliver1998} for a still shorter interval.
Given the visual similarity of AMO and $\Delta T$ in Fig. 2, their 
correlation is surprisingly small, but would increase if the overall upward 
trend of $\Delta T$ were subtracted.
The PDO seems to show no relevant 
correlation with $\Delta T$.

\begin{table}[width=0.99\linewidth,cols=4,pos=h]
\caption{Empirical correlation coefficients $r$ between 
different data sets, each of which represents a centered 
moving average over 23 years. Note the large
values for the correlation both between CO$_2$ and the aa-index 
with $\Delta T$, 
compared to weak or barely existing correlations of the AMO- and PDO-index 
with $\Delta T$.}
\label{tabl1}
\begin{tabular*}{\tblwidth}{@{} LRRR@{} }
\toprule
Correlated data & 1867-2008 & 1867-2003 & 1867-1998\\
\midrule
CO$_2$ with $\Delta T$ & 0.916 &0.894&0.869 \\
aa with $\Delta T$ & 0.806 & 0.900 & 0.950 \\
AMO with $\Delta T$ & 0.260 & 0.164 & 0.106 \\
aa with AMO  & -0.015 & -0.013 & -0.028 \\
PDO with $\Delta T$  & -0.131 & -0.001 & 0.126 \\
aa with PDO & 0.050 & 0.049 & 0.074 \\
\bottomrule
\end{tabular*}
\end{table}

In Fig. 3 we show some further correlation  dependencies, this time
on the time shift $\delta t$ between the two respective data, which possibly 
could give a clue about intrinsic delay effects. One curve 
shows the correlation between the aa-index
and the AMO index, whereby the aa-index at earlier times is correlated 
with AMO at later times.
While the correlation is not large, we see at least a clear maximum 
at $\delta t=11$\,years, as if the AMO-index lagged behind the aa-index by 
this delay time.
The second curve shows the corresponding relationship between $\Delta T$ and
AMO, with the correlation reaching a maximum of 0.3 at $\delta t=6$\,years.
While this looks like a sort of inverted causality (AMO lags behind $\Delta T$)
it could simply mean that $\Delta T$ is indeed governed by the aa-index,
which also determines the AMO-index at later times. 

We also show the corresponding curves for aa-index and $\Delta T$, in the
two versions for the full time interval and the 10-year shortened one.
For the full interval, $r$ starts at the value 0.8 for $\delta t = 0$, but
reaches a maximum of 0.915 for $\delta t = 10$\,years. While on face value
this seems to indicate a 10-years lag  between the aa-index and $\Delta T$,
it is more likely connected with the implied cancellation
of the last 10 years during which the aa-index
was decreasing. In order to test this, 
in the second curve we have omitted
the last 10\,years of both data completely. Here we find an
$r=0.95$ at $\delta t=0$, which still increases to a maximum 
of $r=0.962$ at $\delta t=3$\,yr. This sounds indeed like a reasonable 
time delay between  cause (aa-index) and effect ($\Delta T$).

\begin{figure}[t]		
\includegraphics[width=0.99\linewidth]{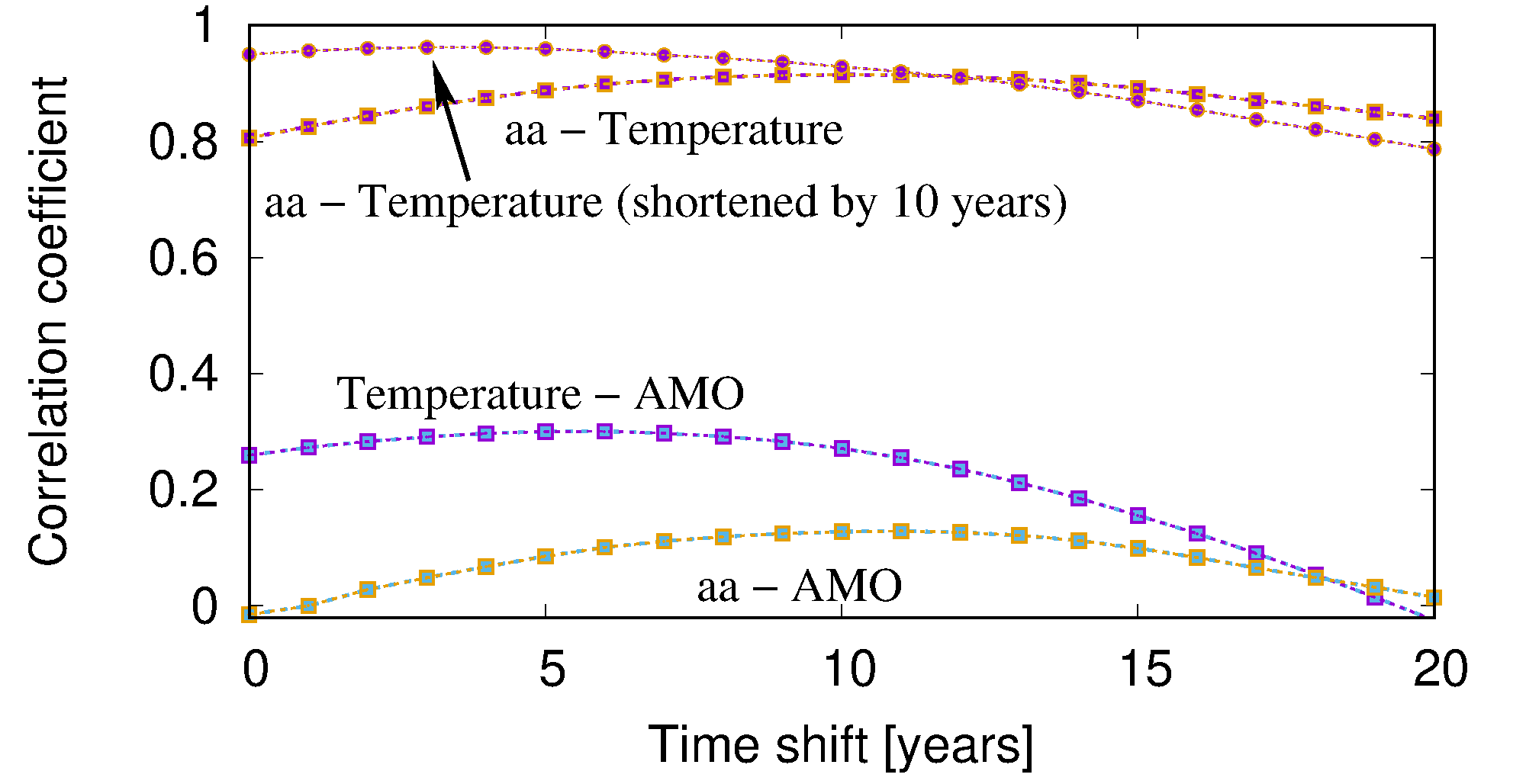}
\caption{Correlations between various data, each representing 
a 23-years centered moving average, in dependence on the time shift $\delta t$
between them. In each case, the first item indicates the earlier data set, the second the
later one. The aa-$\Delta T$ correlation starts at the value $r=0.8$ for $\delta t=0$
and reaches a maximum of $r=0.915$ for $\delta t=10$\,years. While
this might insinuate a 10-years causal time shift between aa-index and $\Delta T$,
it is more likely connected with the canceling effect of the latest decade
with its poor correlation. In the second curve, for which we have completely omitted
the last 10\,years of both data, we get $r=0.95$ at $\delta t=0$, increasing 
to a maximum of $r=0.962$ at $\delta t=3$\,years.
Compared to those large $r$-values, the correlation of  $\Delta T$ and AMO 
is much smaller, viz 0.3 at  $\delta t=6$\,years (AMO lagging behind temperature).
Still smaller is the correlation of aa-index and AMO, with a maximum of
$r=0.129$ for $\delta t=11$\,years. }
\label{FIG:fig3}
\end{figure}

\subsection{Regression}

After those preliminaries, we start now with the multiple regression 
analysis.
For that purpose, we model the temperature data using the 
ansatz 
\begin{equation*}
\Delta T^{\rm model}=w_{\rm aa} \cdot {\rm aa} + w_{\rm{CO_2}} \cdot \log_2({\rm CO_2}/280\, {\rm ppm})
\end{equation*}
with the respective weights $w_{\rm aa}$ and $w_{\rm{CO_2}}$ 
for solar and CO$_2$ forcing,
and compare them with the measured data $\Delta T^{\rm meas}$.
 This procedure is very similar to that of \cite{Soon1996}
who had used though, instead of ${\rm aa}$, 
the length of the sunspot cycle, the averaged sunspot number, and 
a more complicated composite as proxies of solar irradiance.
While \cite{Soon1996} translate all these proxies into
some percentage of  TSI variation (fixing  their  ``stretching 
factor'' by the best fit of modeled and observed temperature 
history), we stick here to the somewhat weird unit K/nT for $w_{\rm aa}$, 
without specifying in detail the physical mechanism(s) underlying the 
solar-climate connection. In case of a dominant Svensmark effect, say, 
this unit would indeed have an intuitive physical meaning, whereas
in case of a dominant UV radiation impact on the ozone layer (plus
subsequent coupling of stratosphere and troposphere), it would just
represent a very indirect, co-responding proxy.

One of the non-trivial questions to address beforehand is 
which time average should be used. Evidently, the
structures of temporal fluctuation of the three data are quite
different. As for the independent data,
the CO$_2$ curve is the smoothest one, so the results of any regression 
will be widely independent on the widths of the 
averaging window. Much more 
fluctuating is the aa-index,
with its dominant 11-year periodicity which had been taken into 
account, though in different ways, by \cite{Cliver1998} and 
 \cite{Mufti2011}.
 \cite{Cliver1998} have been working both with a decadal average
and the so-called aa-baseline ($aa_{\rm min}$), 
i.e. the minimum value which generally 
occurs within one year following the sunspot minima. The empirical
correlation coefficient of $r=0.95$ between $aa_{\rm min}$ and the 
11-year average temperature turned out to be even better than that 
between two decadal averages ($r=0.9$). \cite{Mufti2011}, in turn,  
worked with  11-year and  23-year averages, which will also 
serve as a first guidance for our study,
though later we will consider the dependence on the widths of the
averaging windows in a more quantitative manner.

The dependent variable, i.e. $\Delta T$, is also characterized by
significant fluctuations, although not with the dominant 11-year
periodicity of the aa-index whose climatic impact is thought to be 
smoothed out by the large thermal inertia of the oceans. 
By contrast, $\Delta T$ is strongly influenced by short-term 
variations due to the El Ni\~no-Southern Oscillation (ENSO) 
and volcanism, which had been 
considered in the multiple regression analysis of \cite{Lean2008}.
Our deliberate neglect of those short-term variations, and the focus 
on (multi-)decadal variations, thus requires some averaging on the 
decadal time scale.

In addition to that distinction between different averaging windows, 
we will also consider three cases with different end-years of the 
utilized data. 
In the first case, we take into account all data until 2018. As evident from 
Fig. 1, it is in particular the last decade which shows the strongest 
discrepancy between the decreasing aa-index
and the partly stagnant (``hiatus''), partly increasing temperature 
(in particular during the last El Ni\~no dominated years). Hence, in order 
to asses the specifics of this divergence between the two data, and to 
compare them with the high correlations found by \cite{Cliver1998} and 
\cite{Mufti2011}, we will also consider two shortened periods with 
end years 2013 and 2008, respectively.

Let us start, however, with the full data ending in 2018.
For the two moving average windows (abbreviated henceforth as 
``MAW'') of 11 years and 23 years, Figs. 4a and 4b show the 
{\it fraction of variance unexplained (${\rm FVU}$)}, i.e.
the ratio of the residual sum of squares to
the total sum of squares, in dependence on the
respective weights of the  aa-index ($w_{\rm aa}$, on the abscissa) 
and the  logarithm  of CO$_2$ ($w_{\rm{CO_2}}$, on the ordinate axis).
The latter value provides us immediately 
with a sort of instantaneous climate sensitivity
of the TCR type.
The minimum ${\rm FVU}_{\rm min}=0.107$
is obtained for the weights' combination
$w_{\rm aa}=0.011$\,K/nT and ${\rm w_{CO_2}}=1.72$\,K
in case of  ${\rm MAW}=11$\,years, and 
${\rm FVU}_{\rm min}=0.102$ is obtained for 
$w_{\rm aa}=0.0162$\,K/nT and ${\rm w_{CO_2}}=1.54$\,K
in case of ${\rm MAW}=23$\,years.
From the ellipse-shaped contour plots of 
${\rm FVU}$
we see that those minima reflect a dominating
influence of CO$_2$ over the aa-index.
The two red curves in Fig. 6c show now the 
corresponding temperature reconstructions based on those
optimized values of $w_{\rm aa}$  and ${\rm w_{CO_2}}$, 
for ${\rm MAW}=11$\,years (full line) and for 
${\rm MAW}=23$\,years (dashed line). For both lines we obtain a 
reasonable fit of the general upward trend
of $\Delta T$, but a poor reconstruction of its 
oscillatory features.
This clearly corresponds to the comparable high value
of ${\rm w_{CO_2}}$ compared to that of $w_{\rm aa}$.
Any putative higher share of $w_{\rm aa}$ would lead 
to a drastic decrease of the reconstructed $\Delta T$ for 
the last two decades, resulting in forbiddingly large ${\rm FVU}$ 
values
when compared with the relatively high observed $\Delta T$ 
in this late period.

This brings us to the question of what happens if we
exclude the latest ``hot'' 5 years 
(with their strong El Ni\~no influence), 
thus restricting the date until 2013 only. The corresponding 
results are shown in Fig. 5.
Obviously, the optimal weights' combination now shifts away from CO$_2$
to ${\rm aa}$, with values $w_{\rm aa}=0.0145$\,K/nT and ${\rm w_{CO_2}}=1.56$\,K
for ${\rm MAW}=11$\,years and $w_{\rm aa}=0.0232$\,K/nT and 
${\rm w_{CO_2}}=1.21$\,K
for ${\rm MAW}=23$\,years.
Evidently, the resulting (green) temperature reconstruction
curves in Fig. 5c appear now more oscillatory.

In Fig. 6 we show the corresponding plots for the case 
that we use the end year 2008, which basically corresponds  
to the database of \cite{Mufti2011} 
(2007 in their case).
Evidently, the regression for this shortened segment 
leads to a significantly stronger weight for the aa-index.
The minimum ${\rm FVU}$ is than obtained at
$w_{\rm aa}=0.0190$ K/nT and ${\rm w_{CO_2}}=1.33$\,K
for ${\rm MAW}=11$\,years, and at $w_{\rm aa}=0.0305$\,K/nT and ${\rm w_{CO_2}}=0.80$\,K
for ${\rm MAW}=23$\,years.
Due to the dominance of $w_{\rm aa}$ the (blue) reconstruction curves 
in Fig. 6c (in particular that for  ${\rm MAW}=23$\,years) show
now a significant oscillatory behaviour.

\begin{figure}		
\includegraphics[width=0.99\linewidth]{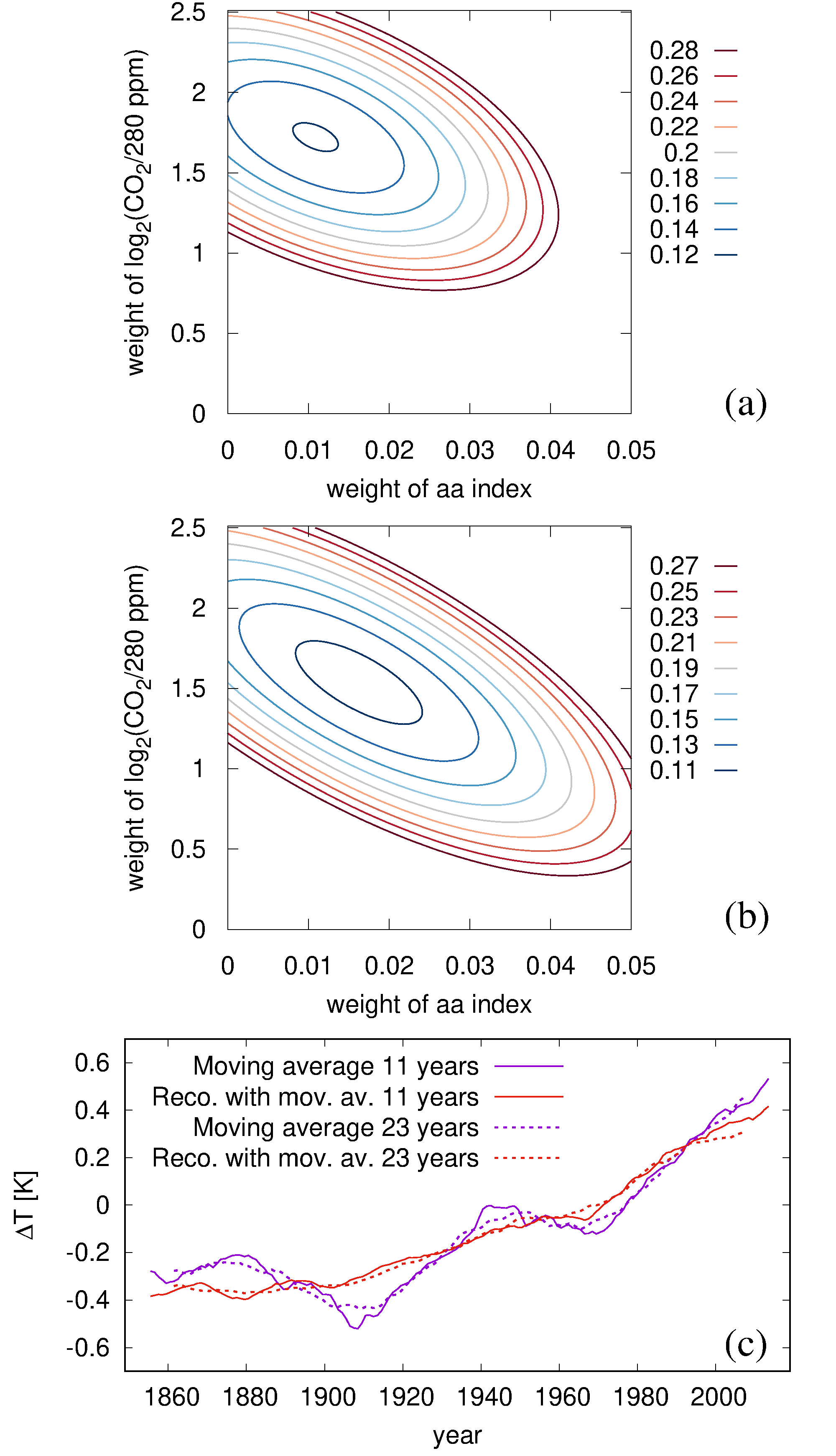}
\caption{Regression analysis for the full data (with end year 2018). (a) 
${\rm FVU}$ in dependence on $w_{\rm aa}$ and ${\rm w_{CO_2}}$ for ${\rm MAW}=11$\,years, with
${\rm FVU}_{\rm min}=0.107$ reached for $w_{\rm aa}=0.011$\,K/nT and ${\rm w_{CO_2}}=1.72$\,K.
(b) The same for ${\rm MAW}=23$\,years, with
${\rm FVU}_{\rm min}=0.102$ reached for $w_{\rm aa}=0.0162$\,K/nT and ${\rm w_{CO_2}}=1.54$\,K.
(c) 11-year and 23-year moving averages for the original $\Delta T$ (purple)
and for the reconstructed $\Delta T$ (red) when using the optimized 
values of $w_{\rm aa}$ and ${\rm w_{CO_2}}$ from (a) and (b).}	
\label{FIG:fig4}
\end{figure}

\begin{figure}		
\includegraphics[width=0.99\linewidth]{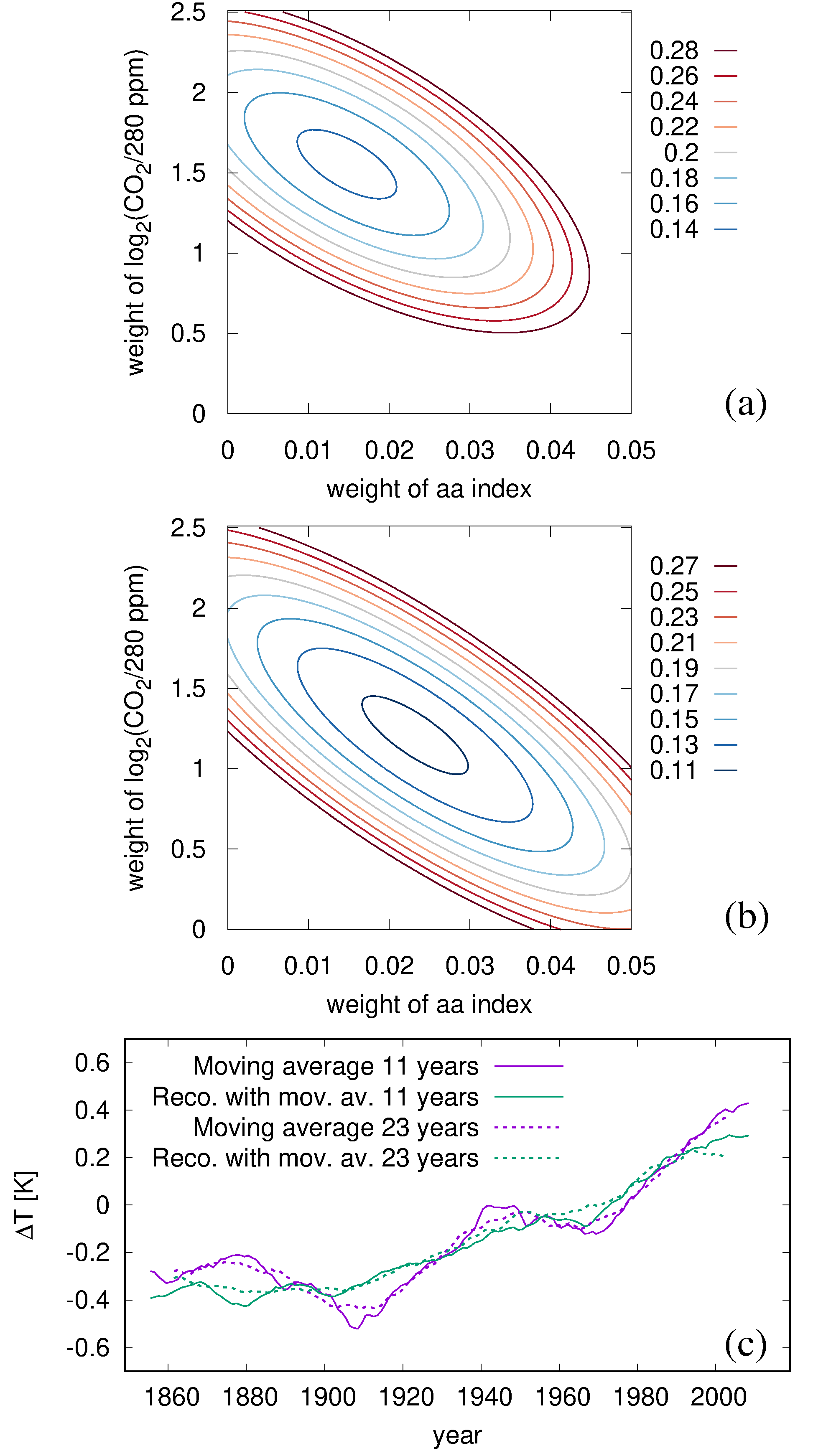}
\caption{Regression analysis for the reduced data (end year 2013). (a) 
${\rm FVU}$ in dependence on $w_{\rm aa}$ and ${\rm w_{CO_2}}$ for ${\rm MAW}=11$\,years, with
${\rm FVU}_{\rm min}=0.134$ reached for $w_{\rm aa}=0.0145$\,K/nT and ${\rm w_{CO_2}}=1.56$\,K.
(b) The same for ${\rm MAW}=23$\,years, with
${\rm FVU}_{\rm min}=0.105$ reached for $w_{\rm aa}=0.0232$\,K/nT and ${\rm w_{CO_2}}=1.21$\,K.
(c) 11-year and 23-year moving averages for the original $\Delta T$ (purple)
and for the reconstructed $\Delta T$ (red) when using the optimized 
values of $w_{\rm aa}$ and ${\rm w_{CO_2}}$ from (a) and (b).}
\label{FIG:fig5}
\end{figure}

\begin{figure}		
\includegraphics[width=0.99\linewidth]{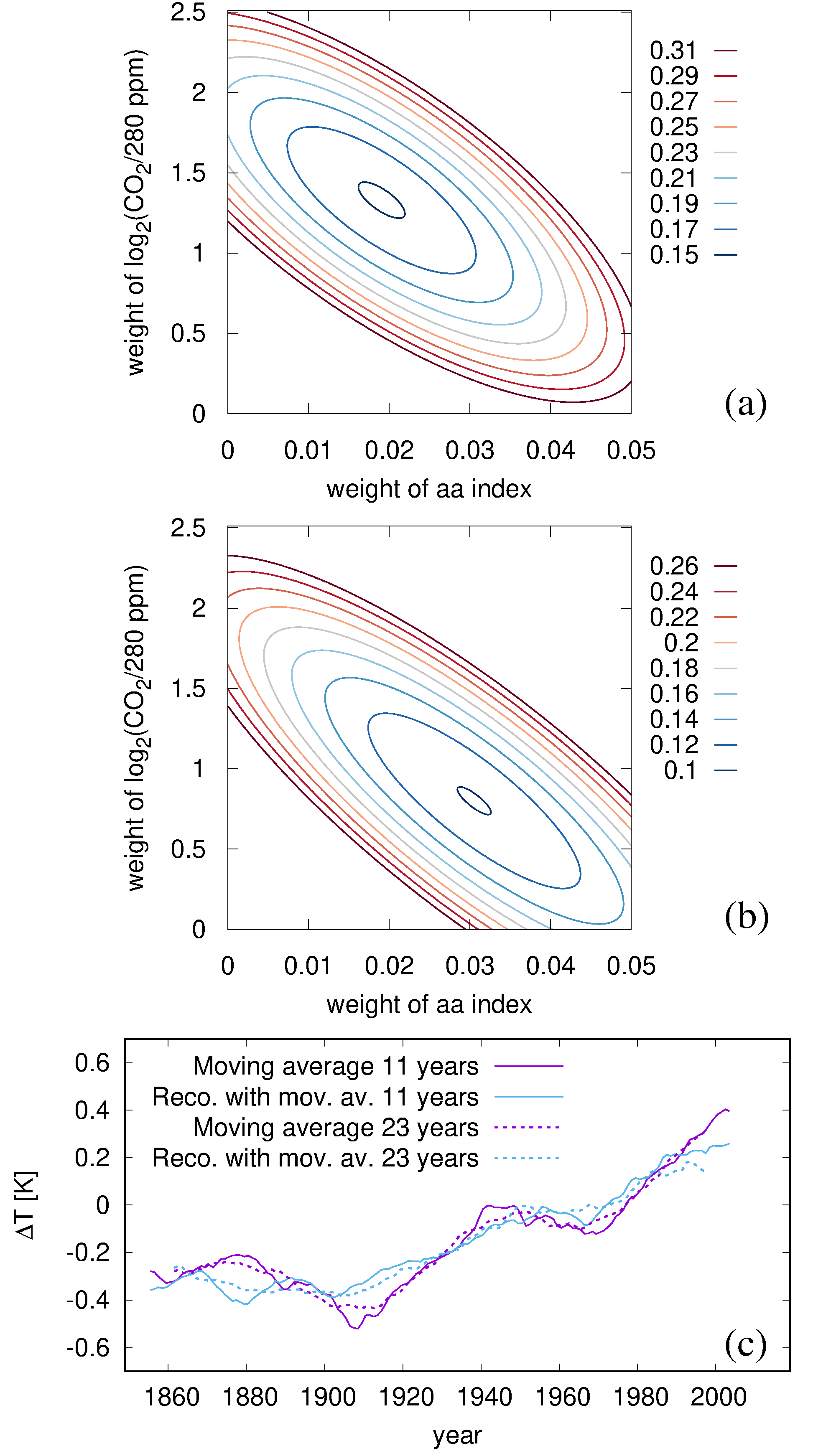}
\caption{Regression analysis for the reduced data (end year 2008). (a) 
${\rm FVU}$ in dependence on $w_{\rm aa}$ and ${\rm w_{CO_2}}$ for ${\rm MAW}=11$\,years, with
${\rm FVU}_{\rm min}=0.149$ reached for $w_{\rm aa}=0.019$\,K/nT and ${\rm w_{CO_2}}=1.33$\,K.
(b) The same for ${\rm MAW}=23$\,years, with
${\rm FVU}_{\rm min}=0.099$ reached for $w_{\rm aa}=0.0305$\,K/nT and ${\rm w_{CO_2}}=0.80$\,K.
(c) 11-year and 23-year moving averages for the original $\Delta T$ (purple)
and for the reconstructed $\Delta T$ (red) when using the optimized 
values of $w_{\rm aa}$ and ${\rm w_{CO_2}}$ from (a) and (b).}
\label{FIG:fig6}
\end{figure}

While those three examples provide a first illustration
of how sensible the solution of the regression reacts 
on the choice  of the end year, and the widths of the ${\rm MAW}$,
the latter dependence will now be studied in more detail.
For that purpose, we analyze first the {\it coefficient
of determination}  $R^2$, which is related to the previously used
${\rm FVU}$ according to $R^2=1-{\rm FVU}$. The corresponding 
dashed curves in Fig. 7a have a rather universal shape,
starting from values of 0.78...0.84 for ${\rm MAW}=3$\,years
to around 0.94  for ${\rm MAW}=39$\,years.
The monotonic increase of $R^2$, which at first glance might suggest 
the use of high values of ${\rm MAW}$, should be treated with caution.
The reason is that a significant share of this increase is just 
due to the increasing ratio of explanatory terms $p$ (in our case two:
$w_{\rm aa}$ and ${\rm w_{CO_2}}$) to the ``honest'' number of data points $n$.
The latter is not identical to the number of considered years, $N_{\rm y}$, 
but - due to the moving average - approximately equal to $N_{y}/{\rm MAW}$
(better estimates, using the lag-one auto-correlation $r^1$ \citep{Love2011},
might still be worthwhile).  
To correct for this effect, we use the so-called adjusted $R^2$, which
is, in general, ${\overline{R}}^2=1-{\rm FVU} \times (n-1)/(n-p-1)$, hence in our 
special case ${\overline{R}}^2=1-{\rm FVU} \times (N_{y}/{\rm MAW}-1)/(N_{y}/{\rm MAW}-3)$.
This adjusted ${\overline{R}}^2$, as a much more telling coefficient of 
determination than $R^2$, is shown with full lines in Fig. 7a.
The monotonic increase still seen for $R^2$ gives now way to a more structured
curve. For the green (data until 2013) and the blue curve (data until 2008)
we observe a local (if shallow) maximum around ${\rm MAW}=25$\,years.
The red curve (data until 2018) has a very flat plateau between
${\rm MAW}=11$ and ${\rm MAW}=27$\,years, with an extremely 
shallow maximum at ${\rm MAW}=25$\,years.
We consider those local maxima
around ${\rm MAW}=25$\,years (${\rm MAW}=27$\,years for the blue curve) 
as a sort of best 
fits. Although ${\overline{R}}^2$ still rises slightly 
for the highest ${\rm MAW}$ values, the corresponding 
number of explained variables becomes then too 
small to allow for a decent reconstruction of the
structure of the $\Delta T$ curve.

The corresponding dependencies for $w_{\rm aa}$ and ${\rm w_{CO_2}}$
are shown in Fig. 7b and 7c, respectively. 
Fig. 7d depicts the same solutions in the two-dimensional
parameter space.
In this representation, all three curves (red, green and blue)
form a sort of common, weakly bent line which appears to connect
the extremal values of ${\rm w_{CO_2}}\approx 1.9$\,K on the ordinate axis
and $w_{\rm aa}\approx 0.04$\,K/nT on the abscissa. This common
line is just a reflection of the 
ellipsoidal shape of the ${\rm FVU}$
as shown in Figs. 3-5 which, in turn, just reflects the
significant ill-posedness of the underlying inverse problem.
To put it differently: due to the high correlation coefficients
of $\Delta T$ both with CO$_2$ as well as with the aa-index, 
the separate  shares of the two ingredients are very hard to 
determine.

However questionable our fixation on the shallow maxima at 
${\rm MAW} \approx 25$\,years
might ever be: Fig. 7 also proves that any alternative choice 
of a higher ${\rm MAW}$ would not change the final solution drastically. 
In either case, both $w_{\rm aa}$ and ${\rm w_{CO_2}}$ converge  
to a close-by value.
Much more decisive is the distinction between the differently colored 
curves: for ${\rm w_{CO_2}}$, they give us results somewhere between
0.6\,K and 1.6\,K (in the latter case we take into account the
particular flatness of the maximum, which in reality might also lay
at slightly lower values of ${\rm MAW}$). 

Actually, this is again a frustratingly wide range of values.
As said, it reflects the ill-posedness of the underlying inverse problem,
whose non-uniqueness leads also to a high sensitivity on 
neglected factors, such as ENSO, AMO, PDO, volcanic aerosols.
At any rate, the temperature development during the next decade 
will be key: if it continues to grow
we will end up at the higher end of the ${\rm w_{CO_2}}$ range. In case 
of some imminent drop it will point to lower values of ${\rm w_{CO_2}}$.

\begin{figure}[h!]		
\includegraphics[width=0.99\linewidth]{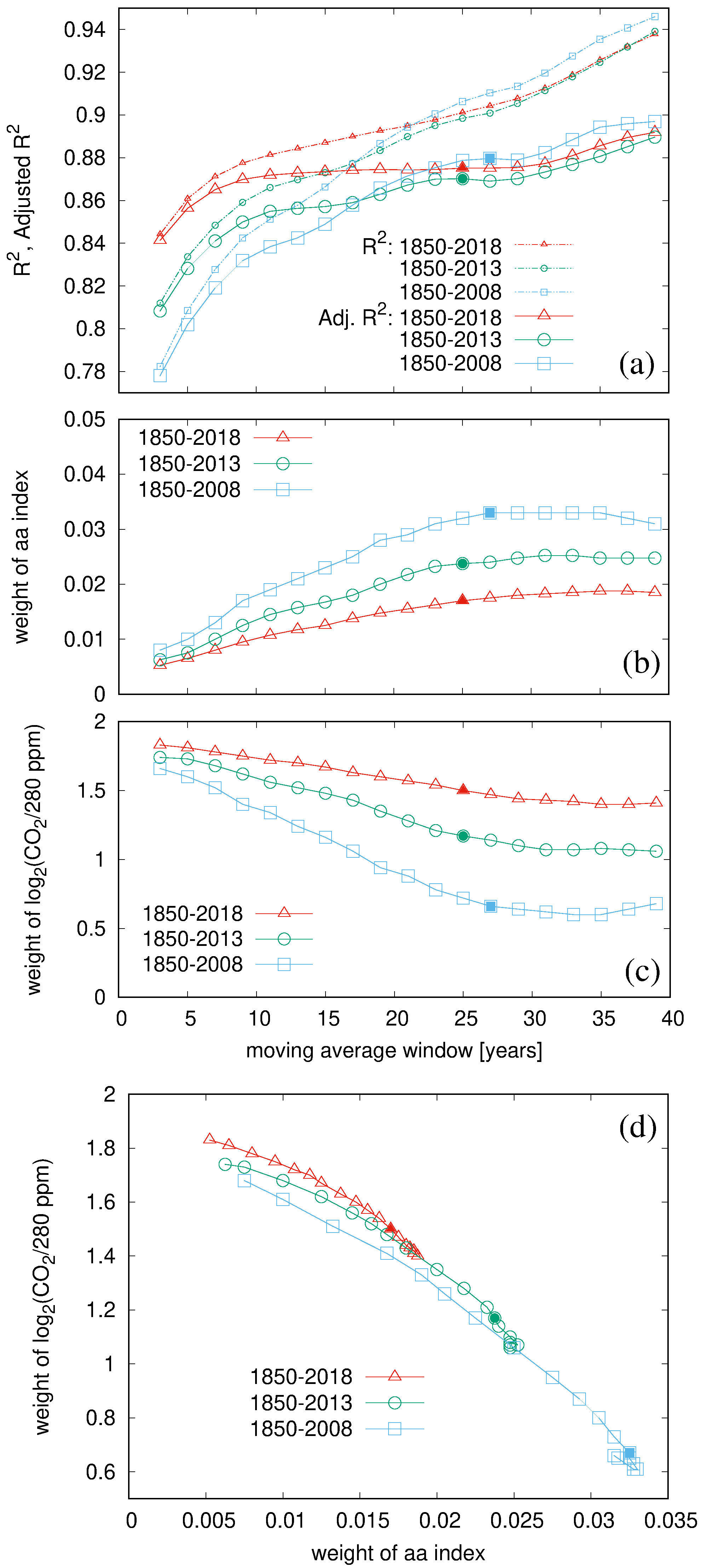}
\caption{Results of the regression in dependence on the width of the moving
average window (${\rm MAW}$). (a)  $R^2$ and its adjusted version  ${\overline{R}}^2$,
each for the three time intervals ending in 2018, 2013, 2008.
The (shallow) local maxima of ${\overline{R}}^2$ 
around ${\rm MAW}=25$\,years are indicated by full symbols.
(b) $w_{\rm aa}$ in dependence on ${\rm MAW}$. The full symbols are the values
corresponding to the local maxima in (a). (c) same as (b), but for ${\rm w_{CO_2}}$.
(d) Regression result in the two-dimensional parameter space 
of $w_{\rm aa}$ and ${\rm w_{CO_2}}$. Note the universal shape of the
solution given a slightly bent, but nearly linear function 
connecting the extremal values ${\rm w_{CO_2}}\approx 1.9$\,K 
on the ordinate axis and $w_{\rm aa}\approx 0.04$\,K/nT on the abscissa.}
\label{FIG:fig7}
\end{figure}

\begin{figure}[h!]		
\includegraphics[width=0.99\linewidth]{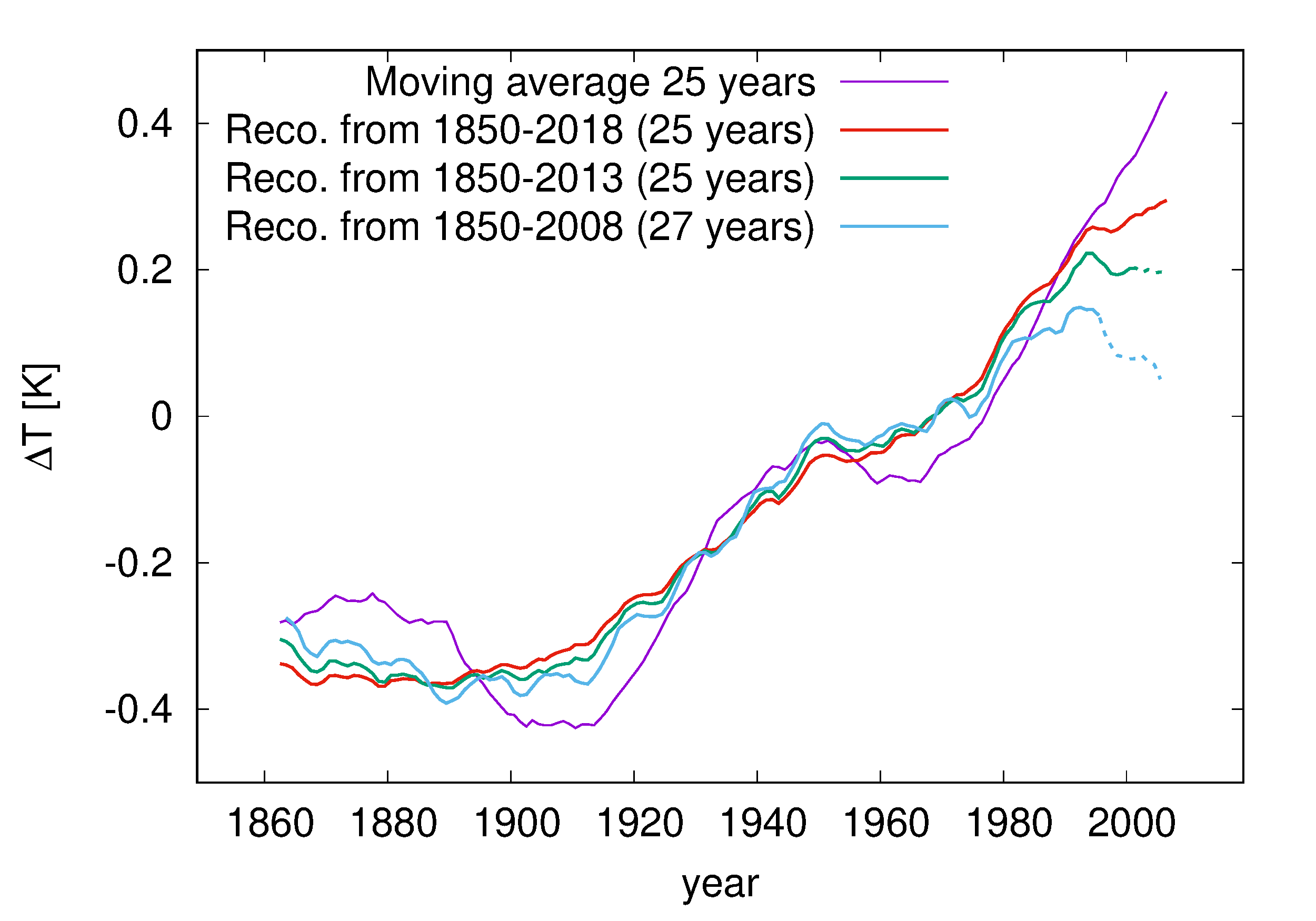}
\caption{Original $\Delta T$ data, averaged over 25 years, and 
reconstructions based on the optimal combinations 
of $w_{\rm aa}$ and ${\rm w_{CO_2}}$
from Fig. 7. The dashed segments of the green and blue curves 
indicate those time intervals that did not enter 
in the respective regressions. From the red over green to blue curve, we 
see an ever improving reconstruction of
the oscillatory behaviour, and an ever increasing divergence with
the observed data at later years.}
\label{FIG:fig8}
\end{figure}

In Fig. 8 we show all three temperature reconstructions
with the optimal combinations of $w_{\rm aa}$ and ${\rm w_{CO_2}}$
taken from Fig. 7, together with the original $\Delta T$ data, averaged 
over 25 years. The dashed segments
of the green and blue curves
correspond to the later time segments which were 
deliberately omitted
in the corresponding regression analysis. As expected, they exhibit an
increasing divergence from the original $\Delta T$ data. On the other
hand, from the red via the green to the blue curve we also observe
an improved reconstruction of the oscillatory behaviour 
of $\Delta T$. This is the crux of our problem: the better
the reconstruction for the years until 1995, say, the larger is
the deviation for the latest two decades. We will need
approximately one decade of more data to being able to 
identify the best solution. For the green or blue lines 
to be valid, a significant temperature drop
in the nearest future will be unavoidable.

\subsection{Some plausibility checks and comparisons}

Before entering the field of predictions, we would like
to check the plausibility of the obtained estimates, in particular those
at the lower end of the range. For that purpose we will shortly discuss
the results of three papers which are based on experimental and 
satellite-borne measurements.

Combining measurements at optically thick samples of CO$_2$
with a 5-layer numerical model for the greenhouse effect, 
\cite{Laubereau2013}
had derived a temperature increase of 0.26\,K for the 
290\,ppm to 385\,ppm increase between 1880 and 2010, which results in
a climate sensitivity of 0.636\,K (per 2$\times$CO$_2$)

\cite{Feldman2015} had published results from two clear-sky surface 
radiative forcing measurements 
with the Atmospheric Emitted Radiance Interferometer (AERI)
between 2000 and 2010, when the CO$_2$ concentration 
increased (according to their estimate) by 22 ppm.
They observed an increase of 0.2 W/m$^2$ during this decade, which amounts to
2.4\,W/m$^2$ (per 2$\times$CO$_2$). With the usual
zero-feedback sensitivity parameter
of 3.7\,K/(W/m$^2$), this translates into 0.65\,K (per 2$\times$CO$_2$). If we were to 
use
the modified value of 3.2\,K/(W/m$^2$) (which
allows for variations with latitude \citep{Soden2006}), 
we get 0.75\,K.

Later on, \cite{Rentsch2019} found a similar result by
analyzing outgoing radiation under night-time, cloud-clear conditions.
The CO$_2$ rise from 373\,ppm to 410\,ppm led to a forcing of 
0.358 W/m$^2$, corresponding to 2.63 W/m$^2$ (per 2 $\times$CO$_2$), 
i.e. to 0.71\,K (with 3.7\,K/(W/m$^2$)) or 0.82\,K 
(with 3.2\,K/(W/m$^2$)),
in nearly perfect agreement with \cite{Feldman2015}.

These three papers are widely consistent among each other, giving
sensitivities in the range between 0.64\,K and 0.82\,K,
which is close to the lower edge of our estimate.
We also note that this end of our regression, which corresponds 
to approximately 70 per cent ``for the sun'', is similar to 
the 50-69 per cent range 
as once found by \cite{Scafetta2007,Scafetta2008}.

This said, we should also note that \cite{Happer2020} have found 
the much higher value of 1.4\,K value (at fixed absolute humidity)
which would fit to our upper limit of 1.6\,K value which is, in turn,
significantly lower than their value 2.2-2.3\,K, 
as inferred for fixed relative humidity. 

The range of our estimates is slightly sharper as the
range 0.4\,K to 2.5\,K of \cite{Soon2015} (with the high value
deemed unrealistic by the authors), and slightly wider than the  
0.8\,K-1.3\,K range of \cite{Lewis2018}, but in either case 
quite consistent with those estimations.

\section{Predictions}

In the preceding section we have derived a certain plausible 
range of combinations of the respective weights of the aa-index and 
the logarithm of CO$_2$ by means of regression analysis 
of data from the past 170 years. 
In the following we will leave the realm of 
solid, data-based science and enter the somewhat 
``magic'' realm 
of predictions.
Given all the underlying uncertainties concerning the future time
dependence of the aa-index, of CO$_2$, and of further climate factors 
such as AMO, PDO, ENSO, volcanism etc., any forecast has to be 
taken with more than one grain of salt. This said, we will at 
least do some parameter studies, by allowing the unknown 
time series of the aa-index and CO$_2$, and their 
respective weights, to vary in some reasonable range. 
Let us start with the aa-index.

\subsection{Predicting the solar dynamo}

This subsection is definitely the most speculative one of 
this paper as it is concerned with forecasts of the aa-index
for the next 130 years.
There is no doubt that the aa-index is strongly correlated 
with the sunspot number (SSN), with typical 
correlation coefficients of
around $r=0.96$ when averaged over one cycle \cite{Cliver1998a}.
Neither is there any doubt that the SSN and the aa-index are both
governed (though in a non-trivial manner) by the solar 
dynamo. So any prediction of the aa-index boils down to 
a prediction of the solar dynamo which many researchers
believe to be impossible, at least beyond the horizon of 
the very next cycle 
for which reasonable (though not undisputed) ``precursor methods'' 
exist  \citep{Svalgaard2005,Petrovay2010}.

\begin{figure}[!h]		
\includegraphics[width=0.99\linewidth]{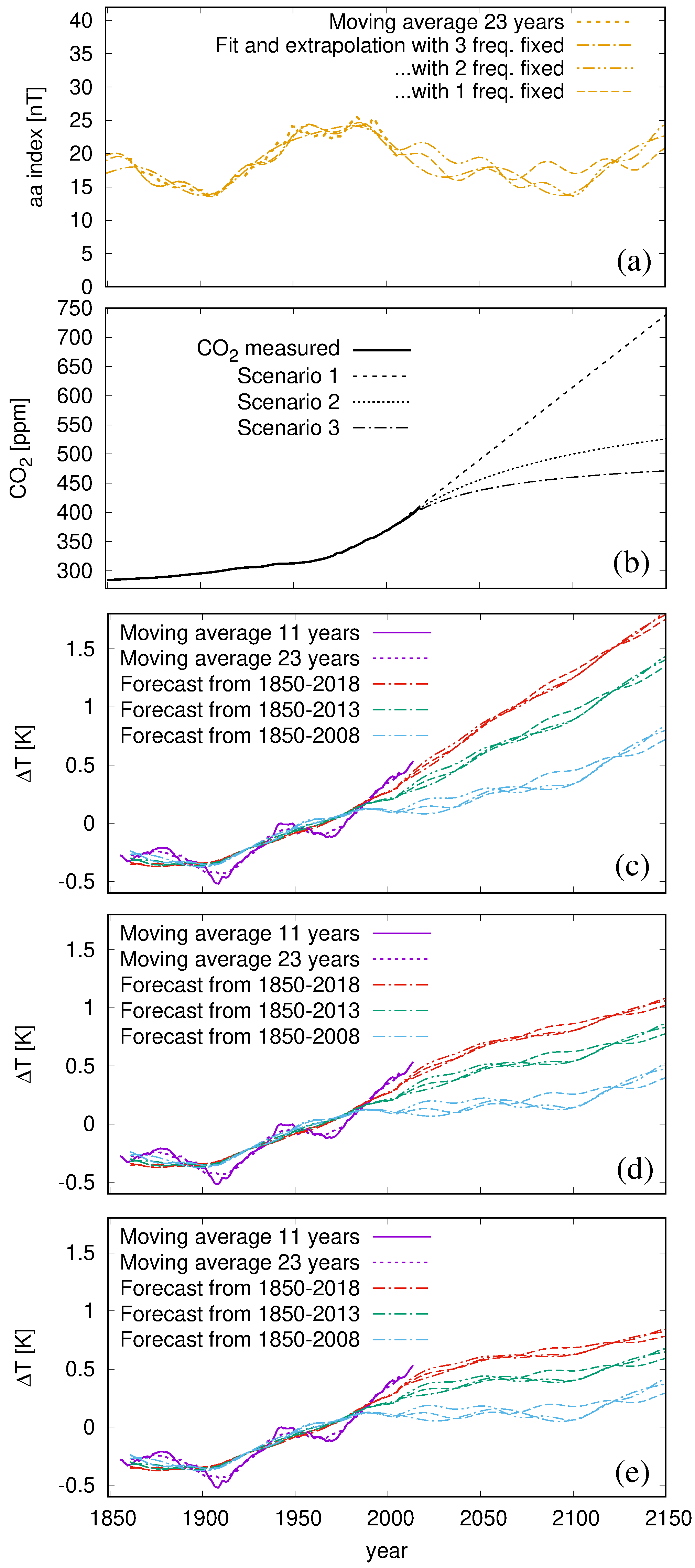}
\caption{Climate predictions until 2150. (a) 23-year moving average of the 
aa-index, and three 3-frequency fits to it, extrapolated  until 2150.
(b) Three scenarios for CO$_2$ concentration.
(c) Temperature forecasts for the 1st CO$_2$ scenario (dashed curve in (b)),
for the three pairs of $w_{\rm aa}$ and ${\rm w_{CO_2}}$ resulting
from regression with end year 2018 (red), 2013 (green), 2008 (blue).
Each of the coloured bundles comprise the three different 
3-frequency fits from (a). (d) Same as (c), but for the second CO$_2$ scenario
with a mild decarbonization scheme
(dotted line in (b)). (e) Same as (c), but for third CO$_2$ scenario
with a radical decarbonization
(dash-dotted line in (b)).}
\label{FIG:fig9}
\end{figure}

In order to justify our audacious forecast for the aa-index, 
we have to make a little diversion on the solar dynamo and its 
short-, medium- and long-term cycles. The reader should be 
warned, though, that our arguments 
do not reflect the main-stream of solar dynamo theory.  We think 
nevertheless that the last years have brought about sufficient 
empirical evidence that justifies at least a cautious try.

Let us start with some recent evidence concerning the 
phase stability of the Schwabe cycle, a matter that 
was first discussed by \cite{Dicke1978}. 
In \cite{Stefani2020b} we have reviewed 
the pertinent results derived from algae data in the early Holocene
\citep{Vos2004} and from sunspot and {\it aurorae borealis} observations, 
combined with $^{14}$C and $^{10}$Be data, from the 
last centuries. Without going into the details
it was shown that the Schwabe cycle is 
very likely phase-stable at least
over some centuries, with a period between
11.04\,years and 11.07\,years. 
While certain nonlinear 
self-synchronization mechanisms
of the solar dynamo cannot be completely ruled out
as an explanation 
\citep{Hoyng1996}, the external synchronization
by the 11.07-years periodic spring tides 
of the (tidally dominant) Venus-Earth-Jupiter system
provides a suspiciously compelling alternative.
Based on previous observations and ideas of 
\cite{Hung2007,Wilson2008,Scafetta2012,Wilson2013,Okhlopkov2016},
we have corroborated 
a model \citep{Weber2015,Stefani2016,Stefani2017,Stefani2018,Stefani2019} 
in which the 
weak tidal forces of Venus, Earth and Jupiter 
serve only as an external trigger
for synchronizing (via parametric resonance) 
the intrinsic helicity oscillations
of the kink-type Tayler instability in the tachocline
region. The arising 11.07-yr period of the helicity parameter
$\alpha$ ultimately leads to the 22.14\,year period of the 
Hale cycle.

Building on this phase coherence of the
Hale cycle, later \citep{Stefani2020a,Stefani2020c}
we exploited
ideas of  \cite{Wilson2013,Solheim2013}
to explain the mid-term Suess-de Vries cycle
as a beat period between the 22.14-yr Hale cycle and the 
19.86-yr synodic cycle of Jupiter and Saturn 
which governs the motion of the Sun around the barycenter
of the solar system. Note that, apart from first ideas 
\citep{Javaraiah2003,Sharp2013,Solheim2013,Wilson2013}, the 
spin-orbit coupling 
that is necessary to translate the orbital motion of the Sun 
into some dynamo-relevant internal forcing, is yet far from understood.
In our model, the Suess-de Vries acquires a clear 
(beat) period of 193 years which is in the lower range of usual 
estimates (but see \cite{Ma2020}). The situation with the Gleissberg
cycle(s) was less clear: those appeared 
as doubled and tripled frequencies of the Suess-de Vries cycle,
but also as independent frequencies resulting from beat periods of 
other synodes of Jovian planets 
with the Schwabe cycle (see Fig. 10 in \cite{Stefani2020c}).
Further below, this vagueness of the Gleissberg cycle(s) will be
factored in when fitting and extrapolating the aa-index.
Lately, in \cite{Stefani2020c} we have tried to explain 
the transitions between regular and irregular intervals 
of the solar dynamo (the ``supermodulation'' as defined by 
\cite{Weiss2016})
in terms of a transient route to chaos.

With those preliminaries, we will now fit the aa-index 
over the last 170 years in order to extrapolate  it
into the future. We assume that we have safely left the irregular 
period of the solar dynamo, as reflected in the Little Ice Age 
which can be considered as the latest link in the
(chaotic) chain of Bond events \citep{Bond2001}. 
Guided by our (double-)synchronization 
model, and encouraged by \cite{Ma2020} who had indeed derived
an 195-yr cycle in the quiet (regular) interval from 800-1340,
we keep this Suess-de Vries period fixed to 193 years in all fits. 
Concerning the Gleissberg-type cycle(s) we will be 
less strict, though. In the first version, we fix the two 
(half and tripled) periods of 96 and 65 years, for the 
second one we only fix the 96 years period, and
for the third one we will keep both Gleissberg cycles 
undetermined.

In the interval between 1850-2150, the results of those three 
different 3-frequency fits 
are presented in Fig. 9a, together with 
the (fitted) 23-year averaged aa-index data between 1850-2007.
Evidently, all three curves approximate the original data 
reasonably well, and all show a similar extrapolation with 
a long decay until 2100, and a recovery afterwards.
This behaviour is quite similar to 
the prediction of \cite{Bucha1998} (their Figure 14) as well as 
with the 
variety of predictions by \cite{Luedecke2015}.
The differences between the three fits mainly concern the 
high frequency part; whereas the version with three frequencies 
being fixed is the 
smoothest, the other two fits comprise some stronger wiggles.
This simply reflects the fact that the Gleissberg cycle 
is more vague than the Suess-de Vries cycle. In the following 
we will always work with all three curves obtained, 
hoping that they constitute a representative 
variety for the future of the aa-index.

\subsection{Some scenarios}

Thus prepared, we consider now three different CO$_2$ scenarios
(Fig. 9b), for each of which we further take into account the three 
predictions for the aa-index as just discussed, as well as 
the three optimal combinations of weights for the aa-index and CO$_2$ as 
derived in the previous section. Deliberately, we restrict our forecasts 
until 2150, admitting that neither the CO$_2$ trend nor
the aa-index are seriously predictable beyond that horizon.

Let us start with the simple case\footnote{While we do 
not refer here to IPCC's Representative Concentration Pathways, 
this case has some similarity to their RCP 6.0, at least until 2100.}
of an unabated linear extrapolation 
of the recent CO$_2$ trend to rise by 2.5 ppm annually (upper curve in 
Fig. 9b), which would bring us to a value 736 ppm in 2150 (still steeper 
trends are not completely ruled out, but perhaps not that realistic 
given the recent worldwide reduction commitments).
For this scenario, Fig. 9c shows three red, green, and blue 
bundles of curves, each comprising the three different fits 
of the aa-index as discussed above. The red bundle (``hot'')
corresponds to the red solution in Fig. 7, with a rather 
high CO$_2$-sensitivity of 1.5\,K and an aa-sensitivity of 0.017\,K/nT.
The green bundle (``medium'') corresponds to the green solution of 
Fig. 7 based on sensitivities of 1.2\,K and 0.022\,K/nT.
The blue bundle (``cool'') corresponds to 0.6\,K and 
0.032\,K/nT.

We see that in 2100 the ``hot'' variant leads to a 
temperature anomaly of 
$\Delta T=1.3$\,K which is $0.9$\,K above the average 
$\Delta T \approx 0.4$\,K
from the first decade of this century. The ``medium'' and ``cool'' 
curves are flatter, leading to $\Delta T=0.9$\,K and
$\Delta T=0.4$\,K, respectively. However, for 
those cases to be of any relevance,
an imminent drop of the temperature would be required to 
reach the corresponding curves before they can continue 
as flat as shown.

The second scenario (dotted line in Fig. 9b) assumes a 
slowly decreasing upward trend of the CO$_2$ concentration 
towards the end 
of the 21st century, with a hypothetical time 
dependence 
\begin{equation*}
{\rm CO_2}=[382+2.5(t-2007)/(1.+(t-2007)/100)]\,{\rm ppm}
\end{equation*}
reaching 
a final value of 529 ppm in 2150. The resulting
red, green and blue curves in Fig. 9d
show a rather flat behaviour.
 
The last (and rather unrealistic) CO$_2$ scenario 
(dash-dotted line of Fig. 9b) assumes a radical 
decarbonization path with 
\begin{equation*}
{\rm CO_2}=[382+2.5(t-2007)/(1.+(t-2007)/50)]\,{\rm ppm} \; . 
\end{equation*}
The resulting temperature curves (Fig. 9e)
are basically constant throughout the end
of our forecast horizon.

\section{Conclusions}

This work has revived the tradition of correlating solar 
magnetic field data  with the terrestrial climate as 
pioneered by \cite{Cliver1998} and \cite{Mufti2011}. Just as 
these authors, we have found 
an empirical correlation coefficient between the aa-index and 
$\Delta T$ with remarkably high values ranging from 0.8 until 0.96, 
which points to a significant influence of solar variability on the
climate. Our modest innovation 
was to employ a multiple (double) regression analysis,
with the logarithm of atmospheric CO$_2$ concentration as the second 
independent variable, whose pre-factor corresponds to an 
(instantaneous)  
climate sensitivity of the TCR type. For a lengths of the
centered moving average window of 25 years 
we have identified optimal parameter combinations 
leading to adjusted $R^2$ values of around 87 per cent. 
Depending on whether to include or not include the data from the
last decade, the regression gave climate sensitivity values from  
0.6\,K up to 1.6\,K (per $2\times$\,CO$_2$), 
and values from 0.032\,K/nT down to 
0.017\,K/nT for the corresponding sensitivity on the aa-index.

Ironically, if interpreted as $1.1 \pm 0.5$\,K, the derived 
climate sensitivity range turns out to have (nearly) the same 
ample
50 per cent error bar as the ``official'' ECS value
($3 \pm 1.5$\,K), which we had criticized in the 
introduction.
Yet, in view of the impressive 95 per cent correlation 
(obtained for the restricted period  1850-2008) we believe 
the correct climate sensitivity 
to be situated somewhere in the lower half of that range. 
This ``bias''
is also supported by the fact that during the last years an 
intervening strong El Ni\~no, 
a  positive PDO, and a positive AMO, have all 
conspired to raise the temperature 
to significantly higher values than what would be
expected from the sole combination of aa-index and CO$_2$. 
With the upcoming switch to La Ni\~na conditions, and
the imminent return of the PDO and AMO into 
their negative phases, we expect  a 
significant temperature drop
for the coming years, which then might re-establish the strong 
correlation between aa-index and $\Delta T$.
At any rate, the next decade will be
decisive for distinguishing between climate sensitivity values
in the lower versus those in the upper half of the derived range.
In this sense we are more optimistic than \cite{Love2011} who believed
that we ``...would have to patiently wait for decades
before enough data could be collected to provide meaningful
tests...''.
Given the high correlation for most of the past 170
years, and the good correspondence with the results 
of recent satellite-borne measurements 
\citep{Feldman2015,Rentsch2019}, our ``bets'' are clearly 
on the lower half of that range.

Based on those estimates, we have also presented some cautious 
climate predictions for the next 130 years. With 
the derived share of the solar influence reaching values between 
30 and 70 per cent, 
such predictions depend critically on a correct forecast of 
the solar dynamo (in addition to that of CO$_2$, of 
course). Following our recent work towards a self-consistent 
planetary synchronization model of short- and medium-term cycles of 
the solar dynamo, we have extrapolated some simple 
3-frequency fits of the aa-index to the data from the last 170 years into 
the next 130 years.
Apart from intrinsic variabilities of such forecasts 
(mainly connected with the vagueness of the Gleissberg-type 
cycle(s)), we
prognosticate  a general decline of the aa-index until 2100, 
which essentially reflects the 200-years Suess-de Vries cycle.
Such a prediction presupposes that we have indeed left 
the irregular solar dynamo episode (corresponding to the Little Ice Age, 
the latest ``Bond event'') and that we will further remain in a regular 
phase of solar activity 
\citep{Weiss2016,Stefani2020c}, similar to that between 
800 and 1340 \citep{Ma2020}.
Of course, we have to ask ourselves whether our prediction 
of a declining aa-index could be completely wrong, with the
Sun eventually becoming even ``hotter'' in the future, thus
adding to the warming of CO$_2$. 
While such a scenario cannot be completely 
ruled out, we consider it as not very
likely, given that the solar activity at the end of the 20th century 
was perhaps the highest during the last 8000 years \citep{Solanki2004},
and that it has declined ever since.
 
As for the CO$_2$ trend, we have considered three scenarios,
comprising an unfettered 2.5\,ppm annual increase until
2150, as well as one soft and one radical decarbonization 
scheme. 
Even in the ``hottest'' case considered, we find only a mild 
additional temperature rise of less than 
1\,K until the end of this century, while all other 
cases result in flatter curves in which the heating effect of increasing 
CO$_2$ is widely compensated by the cooling effect of a decreasing aa-index.
Whatever the rationale of the advocated  $2$\,K goal might be, it 
will likely by maintained even without any drastic decarbonization 
measures.  Apart from that, we also advise that any imminent temperature drop 
(due to the turn of  ENSO, PDO and AMO into 
their respective negative phases)
should not be mistaken as, and extrapolated to, a long-lasting 
downward trend \citep{Abdussamatov2015}.

In this work, we have focused exclusively on  a 
quasi-instantaneous, i.e. TCR-like climate sensitivity 
on CO$_2$. As for ECS, 
we  agree with \cite{Knutti2017} who opined that 
``(k)nowing a fully equilibrated response is of limited value
for near-term projections and mitigation decisions'' and that
``(t)he TCR is more relevant for predicting climate change over 
the next century''.  In view of the 
millennial relaxation time scale underlying the concept 
of ECS, we fear that - perhaps much too soon - the huge Milankovic 
drivers will cool down mankind's hubris of being able to 
significantly influence the terrestrial climate (in whatever 
direction).

\section*{Data availability}
The see surface temperature data HadSST.4.0.0.0 data 
were obtained from 
http://www.metoffice.gov.uk/\\hadobs/hadsst4 on November 27, 
2020 and are $\copyright$ of
British Crown Copyright, Met Office (2020).

The aa-index data between 1868 and 2010 were obtained from
NOAA under
ftp://ftp.ngdc.noaa.gov/STP/\\GEOMAGNETIC$\_$DATA/AASTAR/ .

Monthly aa-index data between 2011 and 2019 were obtained
from the website of the British Geological Survey 
www.geomag.bgs.ac.uk/data$\_$service/data/\\magnetic$\_$indices/aaindex.html .

CO$2$ date were obtained from 
ftp://data.iac.ethz.ch/\\CMIP6/input4MIPs/UoM/GHGConc/CMIP/yr/\\atmos/UoM-CMIP-1-1-0/GHGConc/gr3-GMNHSH/\\v20160701 \; .

\section*{Declaration of competing interest}
The author declares that he has no known competing 
financial interests or personal relationships that could 
have appeared to influence the work reported in this paper.

\section*{Acknowledgment}
This work was supported in frame of the Helmholtz -  RSF  Joint  Research  
Group  "Magnetohydrodynamic  instabilities'',  
contract  No  HRSF-0044. It has  also 
received  funding  from  the European Research  
Council (ERC) under  the European  Union's Horizon 2020 research 
and innovation programme (grant agreement No 787544). 
I'm very grateful to Andr\'e Giesecke, Sebastian L\"uning, 
Willie Soon, Fritz Vahrenholt and Tom Weier for their valuable  
comments on an early draft of this paper.

\end{document}